\documentclass[prb,
twocolumn,
superscriptaddress,showpacs,amsmath,amssymb]{revtex4}

\begin{document}

\title{Negative Newton constant may destroy some conjectures}

\author{G.E.~Volovik}
\affiliation{Low Temperature Laboratory, Aalto University,  P.O. Box 15100, FI-00076 Aalto, Finland}
\affiliation{Landau Institute for Theoretical Physics, acad. Semyonov av., 1a, 142432,
Chernogolovka, Russia}

\date{\today}

\begin{abstract}
The magnitude and sign of the gravitational coupling $1/G$ depend on the relations between different contributions from scalar, fermionic and vector fields. In principle, this may give the zero and negative values of $1/G$ in some hypothetical Universes with the proper relations between the fermionic and bosonic species. We consider different conjectures related to gravity, such as the wave function collapse caused by gravity,
entropic gravity, and maximum force. We find that some of them do not work in the Universes  with zero or negative $1/G$, and thus cannot be considered as universal. Thus the extension of the gravitational coupling to $1/G \leq 0$ provides the test on the universality of the proposed theories of gravity.
\end{abstract}
\pacs{
}

\maketitle

\section{Introduction}

Both in the fundamental gravity and in the induced Sakharov gravity,\cite{Sakharov1968,FFZ1997,Visser2002}
the gravitational coupling  $G^{-1} $ depends on the fluctuating vacuum quantum fields. The contributions of massless fields at zero temperature are (see e.g. Ref.\cite{Broda2009}):
\begin{eqnarray}
G^{-1} = G_0^{-1}+\frac{\Lambda_{\rm uv}^2}{12\pi} \left( n_0 + n_{1/2} -4n_1 \right)\,.
\label{cotunneling}
\label{GzeroT}
\end{eqnarray}
Here $\Lambda_{\rm uv}$ is the UV cut-off  (for the asymptotically safe scenario see review\cite{Pawlowski2021}); $n_0$, $n_{1/2}$ and $n_1$ are the numbers of correspondingly scalar, Weyl and vector fields; and $1/G_0$ contains the other possible contributions, which may include contributions from gravitons, dimension-0 scalars,\cite{Boyle2021,Miller2022} and the "fundamental" value.

For massive fields, the Eq.(\ref{GzeroT}) is modified by the $m^2$ terms, while for the temperature $T$, which is larger than the masses of the fields but smaller than the UV cut-off, the $G^{-1} $ contains the $T^2$ contributions. For example, in effective gravity emerging in the superfluid $^3$He-A with Weyl fermionic quasiparticles, the temperature correction to the inverse Newton "constant" contains the conventional contribution from the Weyl fermions:\cite{Volovik2003,VolovikZelnikov2003} 
 \begin{eqnarray}
G^{-1}(T)  - G^{-1}(T=0) = - \frac{\pi}{18} n_{1/2} T^2\,.
\label{cotunneling}
\label{GNonzeroT}
\end{eqnarray}
Since the thermal contribution does not contain the UV cut-off, it is universal, i.e. it is the same for the fermionic relativistic quantum fields and for the Weyl-like fermions in condensed matter. The same universality takes place for scalar fields: the $(\pi/9)n_0 T^2$ contribution
to $G^{-1}(T)$ comes both from the relativistic scalars in gravity\cite{GusevZelnikov1999}  and from phonons in the effective gravity emerging in superfluid $^4$He.\cite{VolovikZelnikov2003}  

So, we have two separate quantities: the  UV cut-off $\Lambda_{\rm uv}$, which in principle can be fundamental and can be called the Planck energy scale; and the gravitational coupling $1/G$, which cannot be fundamental, since it depends on temperature, species  and their masses. In particular, while $\Lambda^2_{\rm uv}$ is always positive, in the Universes with the proper relations between the fermionic and bosonic fields, the gravitational coupling $1/G$ can be zero and even negative. The negative and zero gravitational  coupling $1/G$ have been discussed in theories with spontaneous breaking of scale invariance by scalar field.\cite{Khuri1982}   
 The possibility of the negative value of $G$ during the cosmological time evolution was also debated, see e.g. Refs.\cite{Starobinskij1981,Ayuso2019}. 
The negative $G$ naturally appears in the effective gravity emerging in the superfluid $^3$He with massless Dirac quasiparticles,\cite{Volovik2020} where it is responsible for the solid angle excess in the effective metric of the global monopole.

The zero value  of $1/G$ may follow from the cancellation of the Weyl anomaly.\cite{Boyle2021} This anomaly cancellation requires conditions 
$n_0=n_{1/2} -4n_1=0$, and if there are no other contributions to $1/G$ in Eq.(\ref{GzeroT}) this leads to $1/G=0$.
The zero value of $1/G$ may also take  place in  the Big Bang, if it is considered as the topological quantum phase transition.\cite{KilnkhamerVolovik2021} In this scenario $1/G$ is always positive, and become zero only at the point of the Big Bang.
The reason for that is that the Big Bang is considered as the intermediate state between two topological vacua with broken conformal symmetry, while in the topologically trivial intermediate state the conformal symmetry is restored, and $1/G=0$.

Some conjectures related to gravity, which are based on the positive value of the gravitational coupling, $G^{-1}>0$, do not work in the Universes with $G^{-1}\leq 0$. That is why such conjectures cannot be considered as universal. Example is provided by the Diosi-Penrose scenario of the collapse of the wave function induced by gravity,\cite{Diosi1984,Diosi1989,Diosi2022,Penrose1996} which is discussed in Sec.\ref{DP}. This scenario does not work in the Universe with $G^{-1} < 0$, and thus cannot be considered as fundamental. Verlinde conjecture of the entropic gravity in Sec. \ref{Verlinde} and the conjecture of the maximum force in Sec.\ref{force} do not work in the Universe with $1/G=0$.

Of course, if $1/G < 0$, the vacuum can be unstable, since the gravitons have negative kinetic energy.
But in a given context, it is important that $1/G < 0$ is, in principle, possible. If some conjecture is fundamental and universal, it should be valid irrespective of the stability or instability of a given state of the quantum vacuum. That is why the extension of the conjecture to the vacua with $1/G \leq 0$ provides the test on the universality of the conjecture.

\section{Diosi-Penrose scenario of gravitationally induced collapse} 
\label{DP}

The physical origin of the collapse of the wave function is a fundamental problem of quantum mechanics. Here we discuss the scenario in which the wave function collapse is caused by gravity.\cite{Diosi1984,Penrose1996}
In the semiclassical approach, which uses the nonlinear Schrödinger equation, there is the gravitational self-interacting term in the energy:
\begin{eqnarray}
U = - \frac{GM^2}{4}\int\int \frac{d^3{\bf r}d^3{\bf r}'}{|{\bf r}-{\bf r}' |}
\nonumber
\\
\left(|\Psi_R({\bf r})|^2-|\Psi_L({\bf r})|^2\right)
\left(|\Psi_R({\bf r}')|^2-|\Psi_L({\bf r}')|^2\right)
 \,.
\label{EnergyDifference}
\end{eqnarray}
Here $M$ is the mass of the body;  $\Psi_R$ and $\Psi_L$ are two wave packets (or in the other similar scenario, the wave functions in the two potential wells, left and right).
For $G>0$ the localization of the wave packet (or the localization of the body in one of the potential wells) is energetically favourable. This according to Penrose \cite{Penrose1996} and Diosi\cite{Diosi1984,Diosi2022} leads to the collapse of the quantum superposition, with the collapse time $\Delta t \sim 1/U$. 

However, for $G<0$ this conjecture does not work: the  quantum superposition of the wave packets has lower energy, thus leading to the "anti-collapse", i.e. the superposition "wins" the game. Since the Diosi-Penrose scenario depends on the sign of the gravitational coupling, and thus on the relations between the quantum fields, it cannot be the fundamental source of the wave function collapse.

The criticism of this scenario of collapse can be found for example  in Ref.  \cite{Adler2007}, where in particular the application of nonlinear Schr\"odinger-Newton equation is criticized, since it violates the linear character of quantum mechanics. 

\section{Verlinde entropic gravity}
\label{Verlinde}
  
Let us first mention that when the entropy of horizon is considered, the contribution of the fluctuating $n_1$ vector fields can be different from the negative contribution of these fields in Eq.(\ref{GzeroT}).\cite{Kabat1995,Donnelly2015,Donnelly2016}  Nevertheless, the negative or zero value of $G^{-1}$ may come from  the negative value of $G_0^{-1}$.

 In the entropic gravity by Verlinde\cite{Verlinde2011} it is assumed that the number of bits of information on the holographic screen 
 is ($\hbar=c=1$):
 \begin{eqnarray}
N= \frac{A}{G} \,.
\label{bits}
\end{eqnarray}
This entropic scenario of emergent gravity does not work in the Universe with $1/G=0$, since the information is simply absent, $N=0$. Nevertheless, in such Universe gravity does exist, since $1/G=0$ simply means that gravitational action starts with the quadratic terms.
Eq.(\ref{bits}) does not make sense for $1/G<0$. All this demonstrates that this entropic approach to gravity is not universal.

If the holographic principle is valid in the Universe with  $G^{-1}\leq 0$, it would be natural if it is determined by the UV cut-off scale: $N \sim A\Lambda^2_{\rm uv}$.  But this has no relation to gravity in the Universe with $G^{-1}<0$, where the Newton's law has the opposite sign.
 
 \section{Jacobson gravity from the first law of thermodynamics}

In the Jacobson conjecture\cite{Jacobson1995} of the thermodynamic origin of Einstein equations (see also review in Ref.\cite{Padmanabhan2010}), the "species problem" (dependence of $G$ on species  and their masses) is taken into account. According to Ref. \cite{Jacobson1994} the Bekenstein-Hawking entropy $A/4G$ contains the same renormalized $1/G$, which comes from the fluctuating  vacuum fields (the possible exception from this rule may come from the vector fields\cite{Kabat1995,Donnelly2015,Donnelly2016}). 

At first glance the extension to the negative $1/G$ looks problematic. This is because the black hole horizons do not exist in the Universe with $1/G<0$. However, in this Universe the de Sitter state can exist if the vacuum energy density $\Lambda$ (the cosmological constant) is negative,  with the Hubble parameter $H^2=(8\pi/3) |G| |\Lambda|$. Thus at least the cosmological horizon at $R=1/H$ may exist, which shows that the local Rindler horizon can be constructed and can be used for the derivation of the Einstein action. 

There are two possible versions for the entropy of the  horizon in the Universe with $1/G<0$: it can be either $S=A/4|G|$, or $S=-A/4|G|$. The negative entropy of horizon was discussed for a white hole obtained from the black hole with the same mass by quantum tunneling,\cite{Volovik2021,Volovik2022} and in Einstein–Gauss–Bonnet gravity, \cite{Odintsov2002,Kruglov2021} see also Ref.\cite{Milton2021} and references therein. It is possible that if there is a meaningful notion of negative entropy, then the Jacobson approach may produce gravity with negative $G$. It would be interesting to consider the connection between the negative $G$ and the negative $S$.

In the Universe with $1/G=0$, the action for gravity consists of the $R^2$ terms, such as in the fourth-order conformal Weyl gravity. In this case the entropy is not proportional to area, and the Jacobson approach should be modfified to obtain the 4-th order gravity.\cite{Jacobson2012}
  
All this would mean that in this approach the entanglement entropy per unit area is not a universal constant, which is the same for all Universes, but  it is the constant which characterizes a given Universe. Note that in the tetrad gravity theory, where the tetrad fields emerge as bilinear combinations of the fermionic fields,\cite{Diakonov2011,VladimirovDiakonov2012,VladimirovDiakonov2014,ObukhovHehl2012,Wetterich2012} the tetrads have dimension of inverse length. Similar dimensional  tetrads emerge in the elasticity theory of solids.\cite{NissinenVolovik2018,NissinenVolovik2019} In this case, the gravitational coupling $1/G$ and the area $A$ become dimensionless.\cite{Volovik2021b} In principle, the dimensionless parameter $1/(4G)$  can be a kind of topological invariant which characterizes the Universes with different topology. Such invariant can take any integer value including the zero and negative values, $1/(4G)=...-2,-1,0,+1,+2, ...$

 \section{Maximal force}
 \label{force}

The maximum force was conjectured by Gibbons,\cite{Gibbons2003} see also recent papers\cite{Faraoni2021,Schiller2021} and references therein:
\begin{eqnarray}
F_{\rm max}= \frac{1}{4G} \,.
\label{MaxForce}
\end{eqnarray}
This is the force between two equal mass static uncharged Schwarzschild black holes
touching each other at the horizon. It was suggested that this is the maximum value of a force between any two objects.
However, in the Universe with $1/G= 0$, any force between the two objects exceeds this limit, and thus this limit is not universal.

On the other hand, in the Universe with $1/G<0$ the black holes are not formed, and thus the force between the gravitational bodies is not limited, i.e. the maximum force does not exist. Also, the cosmic strings  and global monopoles may exist in the Universe with  $1/G<0$. Instead of the angle deficit in metric  outside these defects, there will be correspondingly  the angle excess and solid angle excess.\cite{Volovik2020} As a result the arguments by Gibbons,\cite{Gibbons2003} which are based on the maximum angle deficit, are not applicable.

In principle, it is possible, that there exists the maximum force, which is determined, say, by the UV cut-off,
$F_{\rm max} \sim \Lambda^2_{\rm uv}$. But while this maximum force can be fundamental,  it is not related to the gravitational coupling
$1/G$, which depends on many details and can be made arbitrarily small. Anyway, one should very clearly specify what type of force is considered.\cite{Visser2021}

 \section{Conclusion}

 In our Universe the gravitational coupling $1/G$ is positive. However, there is no rule at all prohibiting the existence of a Universe with a negative or zero value of $1/G$, even if it may be short-lived due to instability. That is why this possibility should be taken into account, when the general principles are introduced. Some conjectures related to gravity do not work in the Universes with $1/G < 0$ or with $1/G = 0$. That is why such conjectures cannot be considered as the universal principles. 
 
The extension of the gravitational coupling to $1/G \leq 0$ provides the test on the universality of the proposed theories of gravity.
In particular, the Diosi-Penrose scenario of gravitationally induced collapse, the Verlinde entropic gravity and maximum force conjecture cannot be extended to $1/G \leq 0$, while the Jacobson approach requires modification. 

It will be interesting to test the other theories,\cite{Bousso2022,Harlow2022} and to exploit the condensed matter systems, where there are different types of emergent gravity (acoustic,\cite{Unruh1981,Unruh2020} from Weyl point,\cite{Volovik2003}  from bilinear combinations of the fermionic fields,\cite{Volovik2021b}  etc.), and different types of event horizons can be simulated.\cite{Volovik2016,Wilczek2020} 
 
  {\bf Acknowledgements}. I thank V. Faraoni, T. Jacobson and A. Zelnikov for discussions and critical comments. This work has been supported by the European Research Council (ERC) under the European Union's Horizon 2020 research and innovation programme (Grant Agreement No. 694248).


\begin{thebibliography}{999}




\bibitem{Sakharov1968} 
A.D. Sakharov,
Vacuum quantum fluctuations in curved space and the theory of gravitation, 
Sov. Phys. Dokl. {\bf 12}, 1040  (1968); Reprinted in Gen. Rel. Grav. {\bf 32}, 365--367 (2000);
Theor. Math. Phys. {\bf 23,} 435 (1976).

\bibitem{FFZ1997} 
V.P. Frolov, D.V. Fursaev and A.I. Zelnikov,
Statistical origin of black hole entropy in induced gravity,
Nuclear Physics B {\bf 486}, 339--352  (1997).

\bibitem{Visser2002} 
M. Visser,
Sakharov's induced gravity: a modern perspective,
Mod. Phys. Lett. A {\bf 17}, 977--992 (2002).

\bibitem{Broda2009}
B. Broda and M. Szanecki,
Induced gravity and gauge interactions revisited,
Phys. Lett. B {\bf 674}, 64--68  (2009),
 arXiv:0809.4203.

\bibitem{Pawlowski2021} 
J.M. Pawlowski and M. Reichert,
Quantum gravity: A fluctuating point of view,
Frontiers in Physics {\bf 8},  551848 (2021).

\bibitem{Boyle2021}
L. Boyle and N. Turok,
Cancelling the vacuum energy and Weyl anomaly in the standard model with dimension-zero scalar fields,
arXiv:2110.06258.

\bibitem{Miller2022}
J. Miller, G.E. Volovik  and M.A. Zubkov,
Fundamental scalar field with zero dimension from anomaly cancelations,
arXiv:2202.05726.
\bibitem{Volovik2003} 
G.E. Volovik, 
{\it The Universe in a Helium Droplet}, 
Clarendon Press,  Oxford (2003).

\bibitem{VolovikZelnikov2003} 
G.E. Volovik and A.I. Zelnikov,  
Universal temperature corrections to the free energy for the gravitational field, 
Pisma ZhETF {\bf 78}, 1271--1276 (2003),
JETP Lett. {\bf 78}, 751--756  (2003);  
gr-qc/0309066. 

\bibitem{GusevZelnikov1999} 
Yu.V. Gusev and A.I. Zelnikov, 
Finite temperature nonlocal effective action for quantum fields in curved space
Phys. Rev. D {\bf 59}, 024002 (1999).

\bibitem{Khuri1982} 
N.N. Khuri,
Sign of the induced gravitational constant,
Phys. Rev. D {\bf 26}, 2664--2670 (1982).

\bibitem{Starobinskij1981} 
A.A. Starobinskij, 
Can the effective gravitational constant become negative?, 
Sov. Astron. Lett. {\bf 7}, 36--38 (1981).

\bibitem{Ayuso2019} 
I. Ayuso, J.P. Mimoso and N.J. Nunes,
What if Newton’s gravitational constant was negative?
Galaxies {\bf 7}, 38  (2019),
arXiv:1903.07604 [gr-qc].

\bibitem{Volovik2020} 
G.E. Volovik,
Vielbein with mixed dimensions and gravitational global monopole in the planar phase of superfluid $^3$He,
Pis’ma v ZhETF {\bf 112},  539--540  (2020),
JETP Lett. {\bf 112},  505--507 (2020),
arXiv:2009.09779.

\bibitem{KilnkhamerVolovik2021} 
 F.R. Klinkhamer and G.E. Volovik,
Big bang as topological quantum phase transition,
 arXiv:2111.07962.
 
\bibitem{Diosi1984}
L. Diosi, 
Gravitation and quantum-mechanical localization of macro-objects, 
Phys. Lett. A {\bf 105}, 199--202 (1984);
arXiv:1412.0201.

\bibitem{Diosi1989}
L. Diosi, 
Models for universal reduction of macroscopic quantum fluctuations,
Phys. Rev. A {\bf 40}, 1165--1174 (1989).

\bibitem{Diosi2022}
L. Diosi, 
On the conjectured gravity-related collapse rate $E_\Delta/\hbar$ of massive quantum superpositions,
AVS Quantum Sci. {\bf 4}, 015605 (2022).

\bibitem{Penrose1996}
R. Penrose,
On gravity's role in quantum state reduction,
General Relativity and Gravitation, {\bf 8}, 581--600 (1996);
R. Penrose, 
Wavefunction collapse as a real gravitational effect, 
in: Mathematical Physics 2000, ed. A. Fokas
et al. (London: Imperial College), 266--282.

\bibitem{Adler2007}
S.L. Adler,
Comments on proposed gravitational modifications of Sch\"odinger dynamics and their experimental
implications,
J. Phys. A: Math. Theor. {\bf 40}, 755--763  (2007). 

\bibitem{Kabat1995} 
D. Kabat,
Nucl. Phys.  B {\bf 453}, 281--299 (1995),  
Black hole entropy and entropy of entanglement 

\bibitem{Donnelly2015} 
W. Donnelly and A.C. Wall,
Entanglement entropy of electromagnetic edge modes,
 Phys. Rev. Lett.  {\bf 114}, 111603 (2015).
 
\bibitem{Donnelly2016} 
W. Donnelly and A.C. Wall,
 Geometric entropy and edge modes of the electromagnetic field,
 Phys. Rev. D {\bf 94}, 104053 (2016).
 
\bibitem{Verlinde2011}
E. Verlinde,
On the origin of gravity and the laws of Newton,
Journal of High Energy Physics. 2011 (4): 29; 
arXiv:1001.0785

\bibitem{Jacobson1995} 
 T. Jacobson, 
 Thermodynamics of spacetime: The Einstein equation of state,
 Phys. Rev. Lett. {\bf 75},1260--1263 (1995),
 arXiv:gr-qc/9504004.
  
\bibitem{Padmanabhan2010}
T. Padmanabhan,
Thermodynamical aspects of gravity: new insights,
Rep. Prog. Phys. {\bf 73}, 046901 (2010).

\bibitem{Jacobson1994} 
T. Jacobson,
Black hole entropy and induced gravity,
arXiv:gr-qc/9404039.

\bibitem{Volovik2021} 
G.E. Volovik,
From black hole to white hole via the intermediate static state,
Modern Physics Letters A {\bf 36}, 2150117  (2021),
arXiv:2103.10954.

\bibitem{Volovik2022} 
G.E. Volovik,
Macroscopic quantum tunneling: from quantum vortices to black holes and Universe,
submitted to JETP issue devoted to Rashba-95,
arXiv:2108.00419.

\bibitem{Odintsov2002}
M. Cvetic, S.Nojiri and S.D. Odintsov,
Black hole thermodynamics and negative entropy in de Sitter and anti-de Sitter Einstein–Gauss–Bonnet
gravity,
Nucl. Phys. B {\bf 628}, 295--330 (2002).

\bibitem{Kruglov2021}
S.I. Kruglov, 
Einstein–Gauss–Bonnet gravity with nonlinear electrodynamics: Entropy, energy emission, quasinormal modes and deflection angle,
Symmetry {\bf 13}, 944 (2021).

\bibitem{Milton2021}
Yang Li, K.A. Milton, P. Parashar and Lujun Hong,
Negativity of the Casimir self-entropy in spherical geometries,
Entropy {\bf 23}, 214 (2021).

\bibitem{Jacobson2012}
R. Guedens, T. Jacobson and S. Sarkar,
Horizon entropy and higher curvature equations of state,
Phys. Rev. D {\bf 85}, 064017 (2012).


\bibitem{Diakonov2011}
D. Diakonov,
Towards lattice-regularized quantum gravity,
arXiv:1109.0091.

 \bibitem{VladimirovDiakonov2012}
A.A. Vladimirov and D. Diakonov,
Phase transitions in spinor quantum gravity on a lattice,
Phys. Rev. D {\bf 86}, 104019 (2012).

 \bibitem{VladimirovDiakonov2014}
A.A. Vladimirov and D. Diakonov,
Diffeomorphism-invariant lattice actions,
Physics of Particles and Nuclei {\bf 45}, 800 (2014).

 \bibitem{ObukhovHehl2012}
Y.N. Obukhov and F.W. Hehl,
Extended Einstein–Cartan theory a la Diakonov: The field equations,
Phys. Lett. B {\bf 713}, 321--325 (2012).

\bibitem{Wetterich2012}
C. Wetterich,
Universality of geometry,
Physics Letters B {\bf 712},  126--131 (2012).

\bibitem{NissinenVolovik2018}
J. Nissinen and G.E. Volovik,
Tetrads in solids: from elasticity theory to topological quantum Hall systems and Weyl fermions,
ZhETF {\bf 154},   1051--1056 (2018),
JETP {\bf 127}, 948--957 (2018),
arXiv:1803.09234.

\bibitem{NissinenVolovik2019}
J. Nissinen and G.E. Volovik,
Elasticity tetrads, mixed axial-gravitational anomalies, and (3+1)-d quantum Hall effect,
PRResearch {\bf 1}, 023007 (2019),
arXiv:1812.03175.

\bibitem{Volovik2021b} 
G.E. Volovik,
Dimensionless physics,
ZhETF {\bf 159}, 815--821 (2021),
JETP {\bf 132},  727--733 (2021),
arXiv:2006.16821.

\bibitem{Gibbons2003}
G.W. Gibbons,
The maximum tension principle in general relativity,
Foundations of Physics {\bf 32}, 1891--1901 (2003).

\bibitem{Faraoni2021}
V. Faraoni,
Maximum force and cosmic censorship,
Phys. Rev. D {\bf 103}, 124010 (2021).


\bibitem{Schiller2021} 
C. Schiller,
Tests for maximum force and maximum power,
Phys. Rev. D {\bf 104}, 124079 (2021).

\bibitem{Visser2021}
A. Jowsey  and M. Visser,
Counterexamples to the maximum force conjecture,
Universe {\bf 7}, 403 (2021).

\bibitem{Bousso2022} 
R. Bousso, Xi Dong, N. Engelhardt, T. Faulkner, T. Hartman, S.H. Shenker and D. Stanford,
Snowmass white paper: Quantum aspects of black holes and the emergence of spacetime,
arXiv:2201.03096.

\bibitem{Harlow2022} 
D. Harlow, B. Heidenreich, M. Reece and T. Rudelius,
The weak gravity conjecture: a review,
arXiv:2201.08380.

\bibitem{Unruh1981}
W. Unruh, 
Experimental black hole evaporation. 
Phys. Rev. Lett. {\bf 46}, 1351--1353 (1981).

\bibitem{Unruh2020}
C. Gooding, S. Biermann, S. Erne, J. Louko, W.G. Unruh, J. Schmiedmayer and S. Weinfurtner,
Interferometric Unruh detectors for Bose-Einstein condensates,
arXiv:2007.07160.

\bibitem{Volovik2016} 
G.E. Volovik,
Black hole and Hawking radiation by type-II Weyl fermions,
Pis'ma ZhETF {\bf 104},  660--661 (2016),
JETP Lett.  {\bf 104},  645--648 (2016),
arXiv:1610.00521.

\bibitem{Wilczek2020}
Y. Kedem, E.J. Bergholtz, and F. Wilczek,
Black and white holes at material junctions,
Phys. Rev. Research {\bf 2,} 043285 (2020),
arXiv:2001.02625.


\end{thebibliography}
\end{document}